\documentclass[12pt]{article}

\usepackage{english,a4}

\usepackage{epsfig}
\usepackage{dcolumn}
\usepackage{amsmath}
\usepackage{cite}
\usepackage{changebar}

\makeatletter \@addtoreset{equation}{section}
\makeatother 
\renewcommand{\thefootnote}{\alph{footnote}}

\newcommand{\scr}[1] {\mbox{\scriptsize #1}}

\newcommand{\DT} {\ensuremath{\Delta(T)\times T^{4}}}

\newcommand{\tc} {\ensuremath{T_{\scr{c}}}}

\begin{document}
\thispagestyle{empty}
\hbox{}
 \mbox{} \hfill \hspace{1.0cm}
         \today 

 \mbox{} \hfill BI-TP 2000/17\hfill

\begin{center}
\vspace*{1.0cm}
\renewcommand{\thefootnote}{\fnsymbol{footnote}}
{\huge \bf  GLUON CONDENSATES
~\\ AT FINITE TEMPERATURE ~\\}

\vspace*{1.0cm}
{\large David E. Miller$^{1,2}$
\\}
\vspace*{1.0cm}
${}^1$ {Fakult\"at f\"ur Physik, Universit\"at Bielefeld, Postfach 100131,\\ 
D-33501 Bielefeld, Germany
\footnote{email: dmiller@physik.uni-bielefeld.de} \\}
${}^2$ {Department of Physics, Pennsylvania State University,
Hazleton Campus,\\
Hazleton, Pennsylvania 18201, USA 
\footnote{permanent address, email: om0@psu.edu} \\}
\vspace*{2cm}
{\large \bf Abstract \\}
\end{center}

We consider various special cases of gluon condensates at finite temperature. 
The gluon condensate for an ideal gas of gluons with a given vacuum expectation
value is introduced for the sake of comparison with that calculated using the 
recent finite temperature lattice gauge simulations for a pure Yang-Mills
$SU(3)$ gauge theory at the known critical temperature. We extend this
comparison using the high precision lattice data for two light dynamical 
quarks. The investigation of these three cases show some interesting 
differences arising from the strong interaction alone and in the presence 
of quarks. In this context we discuss some newer simulations for heavier 
quarks and other properties related to gluon condensation.

\vfill
\noindent PACS numbers:12.38Aw, 11.15Ha, 12.38Mh, 11.40.Dw
\newpage


\section{\bf Introduction}

The present discussions of a possible experimental verification
from heavy ion collisions for a new phase of matter, the Quark-
Gluon Plasma (QGP), bring an added interest in physics at ultrahigh
temperatures. This possibility relates very closely the ideas
of phase transitions with the condensation of gluons and quarks
at finite temperature. 
~\\
\indent 
     In this work we discuss the consequences of the gluon condensate at 
finite temperature for several different special cases.
First we indicate the results of a simple calculation for a gluon gas with 
no interactions having a specified gluon vacuum expectation value as an
ideal gluon condensate where all the interactions are in the ground state. 
Then we use the recent high precision lattice results~\cite{Eng4,Boyd} 
for the equation of state in pure $SU(N_c)$ lattice gauge theory.
The essential relationship for the computation of the gluon condensate 
at finite temperature~\cite{Leut} from the lattice data is the conformal or
trace anomaly which here is due to the scale variance of
quantum chromodynamics (QCD). It relates the trace of the
energy momentum tensor to the square of the gluon field strengths $G^2$
through the renormalization group beta function. Here we shall 
discuss the approach investigated in~\cite{Mill}, for which the
consequences of the finite temperature lattice data for pure $SU(N_{c})$
gauge theory for the gluon condensate~\cite{BoMi} have been presented.
Now, however, we are able to make a comparison of these previous results with
some newer finite temperature simulations for light dynamical quarks
~\cite{MILC97}, which could provide some new insights into the
thermodynamical structure of the QGP.
~\\
\indent
We now briefly discuss the condensation of an ideal gas of gluons, 
which we simply write as $G^2(T)$ for a finite temperature T. 
The vacuum expectation value of the pure gluonic system we denote 
by $G^2_0$, which has a known value at zero temperature~\cite{SVZ1}. 
From dimensional considerations of the structure of the gluon condensate 
we can write down the ideal gluon condensate  
\begin{equation}
  \label{eq:idglucon}
   G^2(T)~=~G^2_0~{\left[1~-~{\left({T \over T_0}\right)}^4 \right]},
\end{equation}
where $T_0$ is the temperature at which the condensation ceases. This equation 
gives a simple relation between the condensate at a finite temperature
T and that at zero temperature. We shall soon relate $G^2_0$ and $T_0$ 
to the corresponding quantities in QCD. Although this form for
$G^2(T)$ as an ideal gluon gas seems presently greatly oversimplified,
earlier analyses in finite temperature QCD had assumed~\cite{BiMi} this
behavior as an approximation to the decondensation of gluons. The extent
to which this assumption is valid we are now able to state. Furthermore,
we will make no attempts at the inclusion of the thermal properties of the
hadrons in the thermodynamics of the gluon condensate. 
~\\
\indent
     We start with a brief discussion of the conformal anomaly at finite
temperature as it relates to the gluon condensate. Thereafter we look
at the lattice data for the cases of pure lattice gauge theory and that with
dynamical quarks. Finally we conclude with a discussion of some newer work
related to these considerations, which further accent the fundamental
differences between these three cases.



\section{\bf The Conformal Anomaly at finite Temperature}

\indent
     The relationship between the trace of the energy momentum
tensor and the gluon condensate has been studied for finite temperature 
by Leutwyler~\cite{Leut} for the investigation of problems in deconfinement 
and chiral symmetry. He starts with a detailed discussion of the 
trace anomaly based on the interaction between Goldstone bosons 
in chiral perturbation theory. Central to his discussion is the 
role of the energy momentum tensor, whose trace is directly related to
the square of the gluon field strength. It is important for us to note that
the temperature dependence of the energy momentum tensor $T^{\mu\nu}(T)$ 
can be separated into the zero temperature or confined part,
 $T^{\mu\nu}_{0}$, and the finite temperature
contribution $\theta^{\mu\nu}(T)$ as follows:
\begin{equation}
  \label{eq:emtensor}
  T^{\mu\nu}(T) = T^{\mu\nu}_{0} + \theta^{\mu\nu}(T) .
\end{equation}
\noindent
The zero temperature part, $T^{\mu\nu}_{0}$, has the standard problems with
infinities of any ground state. 
In what follows we shall not concern ourselves with the confined part
since we are only interested here in the thermal properties. 
The finite temperature part, 
which is zero at $T=0$, is free of such problems. We shall see in the next 
section how the diagonal elements of $\theta^{\mu\nu}(T)$ are calculated in 
a straightforward way on the lattice. The trace  $\theta^{\mu}_{\mu}(T)$ is
connected to  the thermodynamical contribution to the energy density 
$\epsilon(T)$ and pressure $p(T)$ for relativistic fields 
and relativistic hydrodynamics~\cite{LaLi}
\begin{equation}
  \theta^{\mu}_{\mu}(T) = \epsilon(T) - 3p(T) .
  \label{eq:eps-ideal}
\end{equation}
\noindent
The gluon field strength tensor is denoted by
$G^{\mu\nu}_a$, where $a$ is the color index for $SU(N)$.
The basic equation for the relationship between the gluon condensate
and the trace of the energy momentum tensor at finite temperature was
written down by Leutwyler \cite{Leut} using the trace or conformal
anomaly in the following form:
\begin{equation}
\langle G^2 \rangle_{T} = \langle G^2 \rangle_0~-~
\langle {\theta}^{\mu}_{\mu} \rangle_{T}, 
\label{eq:condef}
\end{equation}
\noindent
where the gluon field strength squared summed over the colors is
\begin{equation}
G^{2}~=~{{-\beta(g)}\over{2g^3}} G^{{\mu}{\nu}}_{a}G_{{\mu}{\nu}}^{a}, 
\end{equation}
\noindent
for which the brackets with the subscript $T$ represent thermal expectation
value and the zero stands for the value at zero temperature.
The renormalization group beta function $\beta(g)$ in terms of the
coupling may be written as
\begin{equation}
\beta(g)~=~\mu{dg \over{d\mu}}
         =~-{1 \over{48\pi^{2}}}(11N_{c}~-~2N_{f})g^{3}~+~O(g^{5}).
\label{eq:betafun}
\end{equation}
\noindent
Thus because of the renormalization group beta function $\beta(g)$ the 
quantization of the gluon fields leads to a finite value for the average
of the field strength squared at finite temperature which is related
through the conformal anomaly to the trace of the energy momentum tensor.
The above thermal expectation values for the field strength squared and
the trace of the energy momentum tensor we shall write as $G^2(T)$ and
$\theta^\mu_\mu(T)$ as well as $G^2_0$ for the vacuum expectation value.
 


\section{\bf Gluon Condensation in pure $SU(N_c)$ Lattice Gauge Theory}

The numerical computation of the gluon condensate at finite temperature 
can be carried out in terms of quantities which are evaluated using
lattice gauge theory~\cite{Eng}. Following the usual approach in
in statistical physics we start with a partitition function
${\cal{Z}}(T,V)$ for a given temperature T and spatial volume V.
From this basic quantity we may define the free energy density as follows:
\begin{equation}
\label{eq:frenden}
                   f~=~-{T\over{V}}ln{{\cal{Z}}(T,V)}.
\end{equation} 
The volume V is determined by the lattice size $N_{\sigma}a$, where $a$
is the lattice spacing and $N_{\sigma}$ is the number of lattice steps
in a given spatial direction. The inverse of the temperature T is
similarly determined by $N_{\tau}a$ where $N_{\tau}$ is the number of
steps in the (imaginary) temporal direction. Thus the numerical
simulation is done in a four dimensional Euclidean space with given
lattice sizes $N^3_{\sigma}\times N_{\tau}$, which gives the spatial
volume V as $(N_{\sigma}a)^3$ and the inverse temperature $T^{-1}$ as
$N_{\tau}a$. In an $SU(N_c)$ invariant theory the lattice spacing $a$ is
a function of the bare coupling $\beta$ defined by $2N_{c}/g^2$,
where $g$ is the bare $SU(N_c)$ gauge coupling. Thereby this function
fixes both the temperature and the volume at a given $\beta$. Now we define
the expectation value at zero temperature, $P_{0}$, as well as the spatial and
temporal action expectation values at finite temperature, $P_{\sigma}$ and
$P_{\tau}$, respectively, as the space-space and space-time plaquettes in
the following form:
\begin{equation}
\label{eq:plaque}
P_{\sigma,\tau}~=~1~-~{1\over{N_c}}Re{\langle Tr({U_1}U_{2}{U^{\dagger}_3}
                  {U^{\dagger}_4)}\rangle}
\end{equation}
for the usual Wilson action~\cite{Boyd}. For the symmetric Wilson action
we define the parts ~$S_)$~ as ~$6P_0$~ on the four dimensional Euclidean
symmetrical lattice of size ~$N^4_{\sigma}$~ and ~$S_T$~ as ~$3(P_{\sigma}~
+~P_{\tau})$ on the asymmetrical lattice ~$N^3_{\sigma}\times N_{\tau}$.
Now we may proceed to compute the free energy density by integrating these
expectation values as follows:
\begin{equation}
\label{eq:frenint}
 {f(\beta)\over{T^4}}~=~{{-N^4_{\tau}}{\int\limits^{\beta}_{\beta_{0}} 
                        {d{\beta}^{\prime}}[S_0~-~S_T]},}
\end{equation}
where the lower bound ~$\beta_{0}$~ relates to the constant of normalization.
It is important to note that the free energy density if a fundamental
thermodynamical quantity from which all the other thermodynamical variables
can be gotten.
~\\
\indent
     In order to compute the temperature dependence of the gluon condensate
from the lattice data, we need the dimensionless interaction measure 
$\Delta(T)$~\cite{Eng}. For this reason w define the lattice beta function
in terms of the lattice spacing $a$ and the coupling $g$ as
\begin{equation}
\label{eq:betalat}
\tilde{\beta}(g)~=~-2N_{c}a{dg^{-2}\over{da}}.
\end{equation}
Now we are able to define the interaction measure $\Delta(T)$ as 
\begin{equation}
  \Delta(T)~=~ {N^{4}_{\tau}}{\tilde{\beta}(g)}\left[S_{0}~-~S_{T} \right].
\label{eq:deltacomp}
\end{equation}
\noindent
The crucial part of the recent calculations~\cite{Eng4,Boyd} is the use of 
the full lattice beta function, $\tilde{\beta}(g)$
in obtaining the lattice spacing $a$, or scale of the simulation,
from the coupling $g^{2}$. Without this accurate information
on the temperature scale in lattice units it would not be possible
to make any claims about the behavior of the gluon condensate.
The dimensionless interaction measure is equal to
the thermal ensemble expectation value of $(\epsilon - 3p)/T^4$. Thus by the
equation~\eqref{eq:eps-ideal} above the trace of the temperature dependent 
part of the energy momentum tensor~\cite{Mill,BoMi}, here denoted by 
$\theta^{\mu}_{\mu}(T)$, is equal to the expectation value of $\Delta(T)$
multiplied by the thermodynamical factor of $T^4$, which may be written 
(after suppressing the brakets)as follows:
\begin{equation}
  \theta^{\mu}_{\mu}(T)~=~\DT.
\label{eq:trace}
\end{equation}
\noindent
There are no other contributions to the trace for pure gauge theories on the 
lattice. At zero temperature it has been well understood~\cite{SVZ1}
how in the QCD vacuum the trace of the energy momentum tensor
relates to the gluon field strength squared, $G^{2}_0$.
Since the scale breaking in QCD occurs explicitly at all orders in a loop
expansion, the thermal average of the trace of the energy momentum tensor
should not go to zero above the deconfinement transition.
Thus a finite temperature gluon condensate $G^{2}(T)$
related to the degree of scale breaking at all temperatures, can be defined
from the trace. We have used~\cite{BoMi} the lattice simulations
~\cite{Eng4,Boyd} in order to get the temperature dependent part 
of the trace and, thereby, the value of the condensate at finite temperature.
\begin{figure}[tb]
   \begin{center}
     \leavevmode
     \epsfig{file=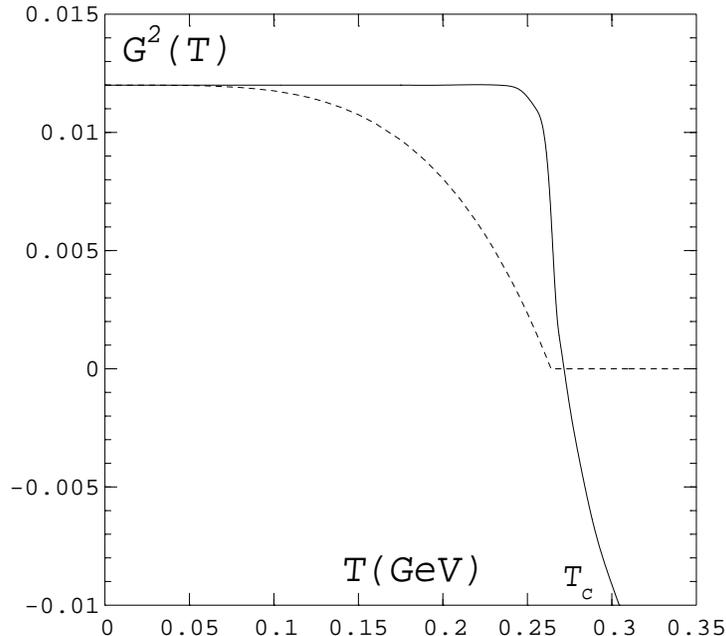
       ,height=90mm
       ,bbllx=85,bblly=240,bburx=484,bbury=611} 
       \caption{The two plots show the gluon condensate for the SU(3) lattice 
gauge theory (solid curve) as compared with that of the ideal gas 
(broken curve). The value for~\tc~ is taken as 0.264 GeV in both cases, above
which the ideal gas condensate is zero.}
       \label{fig:glucon}
   \end{center}
\end{figure}
Accordingly, when we take the gluon condensate~$G^{2}_{0}$~ of the confined 
system as the assumed value~\cite{SVZ1} of~$0.012~GeV^4$~for both the ideal
gluon gas condensate and the pure gauge theory to provide a stable ground 
state. For the ideal gluon gas condensate~$G^{2}(T)$~we use~\eqref{eq:idglucon}
with the condensation temperature~$T_{0}$ equal to~\tc, above which the 
condensate has completely disappeared. 
For the pure $SU(N_c)$ around and above~\tc~we
take the published data~\cite{Eng4,Boyd} for~$\Delta(T)$, and use the
equations~\eqref{eq:condef} and~\eqref{eq:trace} in order to obtain the gluon
condensate~$G^{2}(T)$~as shown in Figure 1. Here we can clearly contrast
the difference between the ideal gluon gas where the condensate has no 
effect at temperatures above~\tc~and the pure SU(3) gauge theory in both
the confined and the deconfined phases. Below~\tc~there is a quantitative
difference between these two cases. Above~\tc~the ideal gas condensate is
completely gone. However, the strongly interacting gluons for~$T~>~T_c$~
arising from the creation of gluons at these high temperatures continue 
the decondensation process into the deconfined phase. Thus we see clearly in
Figure 1 that the pure gluon condensate {\it{never}} approaches an ideal gas 
for all temperatures above~\tc~reached by simulation
~\cite{Eng4,Boyd,Mill,BoMi}.

\section{Gluon Condensation with Light Dynamical Quarks}
 
Now we want to compute the changes in the gluon condensate due to the 
presence of dynamical quarks with relatively small masses.  There have been 
several extensive simulations in recent years of the thermodynamical 
quantities in full QCD with two flavors of staggered quarks~\cite{BKTG,MILC97} 
as well as with four flavors~\cite{BI97}. Although these computations
are still not yet as accurate as those in pure gauge theory, there has
been considerable progress in the actual precision even on not so very large
lattices like $12^3 \times 6$. In the case of the more recent two flavor
simulations~\cite{MILC97} the smaller quark mass shown in Figure 2 with
the open circles  is expressed in lattice units $am_q$ with the value
$0.0125$, which is actually quite small (about~$0.01~GeV$). 
The heavier quark mass of $0.025$ is shown in Figure 2 as the filled squares.
The smaller mass value in terms of the lattice size $a$ has the effect of 
raising the temperature in the simulations so that the lighter quarks are 
at higher temperature for the same coupling. In both these cases we can see
that the general tendency of the temperature dependence of the gluon 
condensation just above the onset of the decondensation shows a sharp
drop which becomes slower at higher temperatures. Because of the scarcity of 
data points above $0.2GeV$ we cannot extrapolate any clear properties 
above this temperature.
\begin{figure}[tb]
   \begin{center}
     \leavevmode
      \epsfig{file=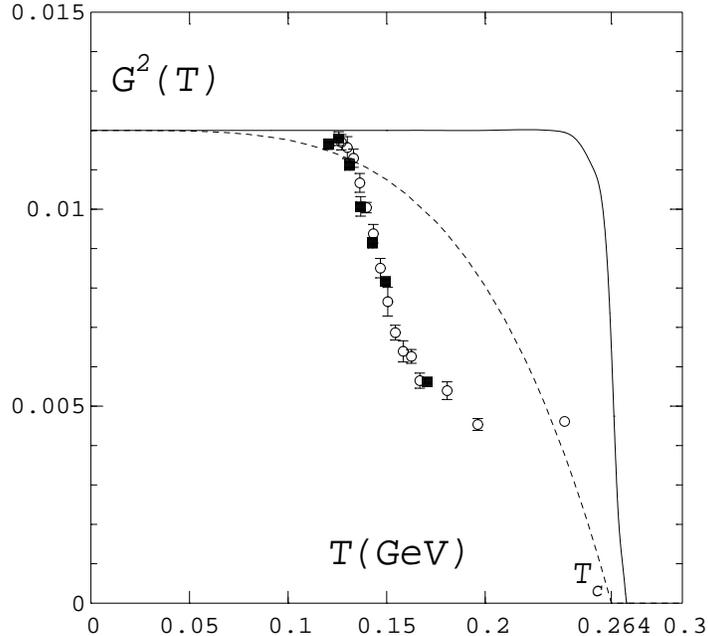,height=90mm,
       bbllx=85,bblly=240,bburx=484,bbury=611} 
     \caption{The lines show the gluon condensates for SU(3) (solid) and the 
ideal gluon gas (broken) in comparison with that of the light dynamical quarks 
denoted by the open circles and the heavier ones  with filled squares. The 
error bars are included when significant.}
     \label{fig:gluconquar}
   \end{center}
\end{figure} 

The presence of massive quarks changes the thermal properties of the
gluon condensate due to the renormalization of the masses as expressed by
the lattice beta function~\cite{BI97}. Thus the trace 
of the energy-momentum tensor has a direct mass dependence in the
trace anomaly \cite{CDJo} for the quark condensate with the renormalized
(physical) quark mass $m_q$. Thereby the expectation value for the 
trace of the energy momentum tensor 
$\langle \Theta^{\mu}_{\mu} \rangle$ is calculated from the sum of the 
expectation values of quark and gluon condensates as 
$m_q\langle \bar{\psi}_q{\psi}_q \rangle~
       +~\langle G^2 \rangle$,
where $\psi_q$ and  $\bar{\psi}_q$ represent the quark and antiquark fields 
respectively. Here for the sake of simplicity we choose two light quarks of 
the same mass. We are now able to write down an
equation for the temperature dependence of the gluon condensate~\cite{Mill} 
including the effects of the light quarks in the trace anomaly in the 
following form:
\begin{equation}
\langle G^2 \rangle_T ~=~ \langle G^2 \rangle_0
~+~m_q\langle\bar{\psi}_q{\psi}_q \rangle_0~
~-~m_q\langle\bar{\psi}_q{\psi}_q\rangle_T~
~-~\langle \Theta^{\mu}_{\mu}\rangle_T.
\label{eq:fullcond}
\end{equation}
As we have discussed previously, we identify the expectation value at finite 
temperatures for the gluon condensate with $G^2(T)$ in Figure 2. 
The simulation procedure is, in general, similar to the pure gauge theory 
with the additional effects of the quark contributions. 
The computation~\cite{MILC97} of the interaction
measure $\Delta_{m}(T)$ where the subscript m denotes its dependence upon
the quark mass $m_q$. As before, we can calculate the temperature dependence
of the trace of the energy momentum tensor (omitting the brackets)
\begin{equation}
\Theta^{\mu}_{\mu}(T)~=~\Delta_{m}(T)\times T^4 .
\end{equation}
This high precision data for $\Delta_{m}(T)$ is used for the compuatation
of the two sets of data points in Figure 2 mentioned above~\cite{MILC97}.
It is possible to see that in equation~\eqref{eq:fullcond} at very low 
temperatures the additional contribution to the temperature dependence 
from the quark condensate with small quark masses is rather insignificant 
as would be expected from chiral perturbation theory~\cite{Leut}. 
However, in the temperature range where the chiral symmetry
is being restored there is an additional effect from the term
$\langle \bar{\psi}_q{\psi}_q\rangle_T$. Well above \tc\/ after the
chiral symmetry has been completely restored the only remaining
effect of the quark condensate is that of the vacuum. This term would 
contribute negatively to the gluon condensate. Thus we expect~\cite{BoMi} 
that for the light quarks the temperature dependence can only contribute 
at all around \tc\/. Nevertheless, because of the small quark mass the total 
effect of the light quark condensate is very small
for the above mentioned simulated~\cite{MILC97} values (maximally
less than one procent). 
For the case of pure $SU(N)$ we know that the values just below \tc\/
of \DT\/ are very small~--~that is, about the same size as
$m_q\langle \bar{\psi}_q{\psi}_q \rangle_0$. Thus, to the extent that
the chiral symmetry has not been completely restored, its effect on 
$G^{2}(T)$ will be present but small just below \tc\/.
~\\
\indent
Finally as a conclusion to this discussion of the gluon condensate with
two light dynamical quarks we shall mention a few related points.
While for the simulations of the pure $SU(N_{c})$
gauge theory we could depend on considerable precision in the determination
of \tc\/ and $\Delta(T)$ as well as the numerous other thermodynamical 
functions, this is still not the case for the theory with the massive 
dynamical quarks. The statistics for the 
measurements are generally smaller. The determination of the temperature
scale is thereby hindered so that it is harder to clearly specify a given
quantity in terms of $T$. Thus, in general, we may state that
the accuracy as well as the number of data points for the case of the
dynamical quarks are much less as compared with the computations  
of the pure lattice gauge theories. However, there is a point that arises
from the effect that the transition temperatures with massive quarks are 
generally  considerably lower, so that \DT\/ is much smaller~\cite{BoMi}. %
Nevertheless, there could be an indication of how the stability of the 
simulations with light dynamical quarks keeps $ G^2(T)$ in positive values 
for $T~>~\tc$ so that amount of decondensation is less than the value
$0.012GeV^4$~\cite{SVZ1} in the temperature range of the numerical data
~\cite{MILC97}. A model of the thermodynamics of the chiral restoration
transition~\cite{kochbrown} has previously considered separately the 
condensates for the electric and magnetic field strengths squared. A
general qualitative agreement of the local subtraction of the magnetic and
electric condensates can be seen with our results in Figure 2 even 
though the lattice measurements which they used were not computed
using a non-perturbative method~\cite{Eng3}, nor was the temperature scale
obtained from the full non-perturbative beta-function~\cite{Eng4,Boyd}. 



\section{Discussions and Conclusions}

In this work we have compared the effects of gluon condensation
at finite temperature for three different cases-- that of 
noninteracting gluons, the pure $SU(3)$ thermodynamics and QCD
with two light quarks. We have seen that the ideal gluon gas has 
the typical structure of nonrelativistic Bose-Einstein condensation,
whereby the condensate completely vanishes at a given temperature 
$T_{0}$ and is exactly zero at all higher temperatures. For the
condensate at zero temperature $G^2_0$ we use the known vacuum expectation
value~\cite{SVZ1}. This behavior we have contrasted with that for the
decondensating of gluons found from the pure $SU(3)$ lattice data,
for which the decondensation does not stop at the critical temperature
but continues as the temperature increases since there is no bound
on the number of thermal gluons~\cite{BoMi}. The presence of two light
dynamical quarks further changes the process of decondensation by 
providing other thermal mechanisms through which the gluon condensate
begins to disappear at a much lower temperature. Although we have seen
that this effect is slightly dependent on the value of the mass, it is 
strongly dependent on the fermionic structure relating to the chiral
symmetry restoration at temperatures lower than the critical temperature~\tc~
of the pure $SU(N_c)$. Thus we are seeing the thermal effects of
two different anomalies~\cite{Mill1} at work on the same system.
~\\
\indent
     We now mention some other work which is relevant to our discussion.
The very recent calculations performed for heavier quarks~\cite{KLP} provide
some new aspects to the above analysis. At the present the pressure has been 
computed as a function of the temperature up to about $4T_c$ for flavor 
numbers 2, $2+1$ and 3, where $2+1$ means two lighter and one heavier 
quark flavors. The simulations present for other quantities considerable 
difficulty with the 
$m_q \langle \bar{\psi}_q {\psi}_q \rangle_T$ term when 
$m_q$ is larger as is the case for the s-quark. 
Furthermore, there has been some recent work on the temperature
dependence of the gluon condensate derived from the QCD sum rules~\cite{HGF}.  
One can see from their results pronounced differences in the thermal behavior 
between this method and our calculation using the lattice simulations
~\cite{Mill,BoMi}. In fact in some cases they find that the ratio 
$G^2(T)/G^2_0$
is larger than unity and growing for increasing temperatures.
As a last discussion we would like to comment on some results based on the
postulation of a simple form for the entropy~\cite{AsHa}. With the assumption
of the proper parameters relating to a bag type of model they calculate 
a form of $\Delta(T)$ which can lead to a similar shape of the
$G^2(T)$ for the pure gauge theory shown in Figure 1. However, all of the
parameters need to be determined from the lattice simulation or the
phenomenological models independently.

\medskip
\noindent{\bf\Large Acknowledgements}

\medskip
The author would like to thank Rudolf Baier, Frithjof Karsch, Reinhart
K\"ogerler, Edwin Laermann and Helmut Satz for many very helpful discussions. 
A special thanks is given to Graham Boyd with whom much of the earlier work 
was carried out. He is very grateful to J\"urgen Engels for the use of his 
programs and many valuable explanations of the lattice results and 
thanks Carleton DeTar for sending the MILC two light flavor data. 
                                                                     


\end{document}